\documentclass[namedreferences]{SolarPhysics}
\usepackage[optionalrh]{spr-sola-addons} % For Solar Physics 
\usepackage{graphicx}                    % For eps figures, newer & more powerfull
\usepackage{color}                       % For color text: \color command
\usepackage{url}                         % For breaking URLs easily trough lines
\usepackage{solaheader}	%	Remove the leading % to remove the copyright statements
\begin{document}

\begin{article}

\begin{opening}

\title{Automatic Detection of Limb Prominences in 304 \AA\ EUV Images}

%%%%%%%%%%%%%%%%%%%%%%%%%%%%%%%%%%%%%%%%%%%%%%%%%%%
%% Authors Names
%
\author{N.~\surname{Labrosse}$^{1}$\sep
        S.~\surname{Dalla}$^{2}$\sep
        S.~\surname{Marshall}$^{3}$      
       }

%%%%%%%%%%%%%%%%%%%%%%%%%%%%%%%%%%%%%%%%%%%%%%%%%%%
%% Runningheads
%
%\runningauthor{}
%\runningtitle{}

%%%%%%%%%%%%%%%%%%%%%%%%%%%%%%%%%%%%%%%%%%%%%%%%%%%
%% Affilations 
%
  \institute{$^{1}$ Department of Physics and Astronomy, University of Glasgow, Glasgow G12 8QQ, Scotland
                     email: \url{n.labrosse@physics.gla.ac.uk}\\ 
             $^{2}$ Jeremiah Horrocks Institute for Astrophysics and Supercomputing, University of Central Lancashire, England\\
	     $^{3}$ Department of Electronic and Electrical Engineering, University of Strathclyde, Scotland\\
%                     email: \url{e.mail-c} \\
             }

%%%%%%%%%%%%%%%%%%%%%%%%%%%%%%%%%%%%%%%%%%%%%%%%%%%
%%% Abstract 
\begin{abstract}
	A new algorithm for automatic detection of prominences on the solar limb in 304~\AA\ EUV images is presented, and results of its application to SOHO/EIT data discussed. The detection is based on the method of moments combined with a  classifier analysis aimed at discriminating between limb prominences, active regions, and the quiet corona.  
	This classifier analysis is based on a Support Vector Machine (SVM). Using a set of 12 moments of the radial intensity profiles, the algorithm performs well in discriminating between the above three categories of limb structures, with a misclassification rate of 7\%.
	Pixels detected as belonging to a prominence are then used as starting point to reconstruct the whole prominence by morphological image processing techniques.
	 It is planned that a catalogue of limb prominences identified in SOHO and STEREO data using this method will be made publicly available to the scientific community.
\end{abstract}

%%%%%%%%%%%%%%%%%%%%%%%%%%%%%%%%%%%%%%%%%%%%%%%%%%%
%% Keywords
%

\end{opening}

\section{Introduction} \label{s:intro} 

Solar prominences are large structures confining a cool and dense plasma in the hot, tenuous solar corona. Once they are formed, they may remain relatively stable for long periods, over a solar rotation or more. The exact processes taking place during formation, evolution, and disappearance are, however, not well understood.
Most studies dedicated to solar prominence physics so far have dealt with individual examples of well-observed structures, and there have been very few attempts to make large statistical studies of their properties \cite[for example]{gilbertetal00,2008SoPh..248...51M}. There are still large uncertainties over prominence physical parameters, and it is not clear how some plasma properties may vary from one prominence to one other (see the review by \opencite{2002SoPh..208..253P}).

Understanding the cause of the disappearance of prominences is essential in many respects. Prominences usually disappear during an eruption where a great amount of matter and energy is released within a few hours. These dynamic events are sometimes linked to coronal mass ejections (CMEs) and/or flares. CMEs and flares are a major source of disturbances in the interplanetary medium, which is of course critical for human activities in space, including around the Earth. Erupting filaments (seen on the disk from the Earth) can be at the origin of Earth-directed CMEs and flares.
One major issue faces those trying to understand prominence properties and to unveil the mechanisms leading to the eruptions: the search for a large number of relevant observations. 
At present, no catalogue of off-limb solar prominences is available to scientists: our primary objective is to generate such a catalogue and to make it available within the Virtual Observatory.
In order to create a catalogue of solar prominences and facilitate this task, we adopt the following principles:
\begin{itemize}
	\item We use data from spacecraft that perform near-continuous observations of the Sun in order to make a list of as many events as possible. We want to use full-Sun images in order to catch all events at any one time. The best instruments corresponding to this description are so far the \textit{Extreme Ultraviolet Imager} (EIT) on the \textit{Solar and Heliospheric Observatory} (SOHO), and the EUVI on STEREO/SECCHI. These datasets are publicly available. The AIA imaging suite on SDO will also match these criteria once launched and in operation. 
	\item Prominences are best observed from space in the resonance line of He\,{\sc ii} at 304~\AA. This line is (or will be) observed by the three instruments mentioned above (EIT, EUVI, AIA). EIT and EUVI also have other channels corresponding to a higher temperature emitting plasma (typically the coronal plasma). In these channels, prominences are usually seen as dark features, because they absorb the coronal radiation from behind them, and because the cool prominence plasma is not hot enough to emit in these lines \cite{ah05,2008ApJ...686.1383H}. AIA on SDO will also have  additional channels where prominences and filaments are expected to be seen in absorption.
\end{itemize}
In this work we will be looking at off-limb structures only. The detection of structures on the disk is deferred to a later study.
The automatic detection of solar prominences and the ensuing construction of a solar prominence catalogue will provide  researchers with interests in understanding the formation, stability, and disappearance of prominences and filaments with a new tool. It will also be valuable for studying the links between the different manifestations of large-scale eruptions on the Sun: eruptive prominences and filaments, CMEs, flares. It will be possible to link the prominence catalogue with filament \cite{2005SoPh..228..361Z} and flare \cite{2009SoPh..256....3A} and CME \cite{2009EM&P..104..295G,2009ApJ...691.1222R} catalogues. Having access to a large statistical sample of these events will make it easier to check theories against observations. In addition to solar prominences, our algorithm identifies active regions on the limb, and this information will also be compiled into a catalogue.

In recent years, there has been considerable interest in developing methods for automated detection of prominences.
The majority of efforts focussed on the detection of prominences on the solar disk, \textit{i.e.}~ filaments.
\inlinecite{2005SoPh..228...97B} used a method based on thresholds and morphological filtering to detect filaments in H$\alpha$ images. The EGSO Solar Feature Catalogue included filament information, obtained from H$\alpha$ data by means of image enhancement and thresholding followed by a region-growing procedure \cite{2005SoPh..227...61F}. Issues related to the tracking of filaments over consecutive solar images have been discussed by \inlinecite{2008Aboudarham}. Detection of filaments using EUV data has been achieved using the four wavelengths of SOHO/EIT complemented by magnetogram data, by means of an image segmentation technique using  a combination of region-based and edge-based methods \cite{2008SoPh..248..425S}.

As far as prominences on the limb are concerned, \inlinecite{2006SoPh..234..135F} considered SOHO/EIT images at 171, 195, 284, and 304~\AA, and applied multiple techniques, including bright histogram segmentation and edge detection to automatically identify prominences, making use of the fact that active regions contribute to emission in all four wavelengths, while prominences appear bright only in the 304~\AA\ images. Other studies focussed on the detection of limb prominence {\it activity} (\textit{e.g.}~eruptions) using data from the Nobeyama Radioheliograph \cite{2006PASJ...58...85S} and H$\alpha$ images from Big Bear Solar Observatory \cite{2007Fu}.

The technique of \inlinecite{2006SoPh..234..135F} suffers from the necessity of having a complete set of images taken at various wavelengths in order to find the off-limb prominences. Here we are interested in limiting the detection of off-limb structures to one wavelength only. The main argument for this is that this method will be more widely applicable to a range of instruments as long as they observe in this particular wavelength (here 304~\AA). The success will not depend on the availability of other types of observations.
There are also issues regarding processing speed, which are crucial if we want to apply this technique for \textit{on the fly} prominence detection on high-cadence SDO images. 

This paper presents a new method for detection of limb prominences, based on moments of radial intensity profiles in 304 \AA\ images. The method has so far been applied to SOHO/EIT images, and its extension to STEREO/EUVI and SDO/AIA data is envisaged in the near future.
In Section~\ref{s:profiles} we describe the intensity profiles that are extracted from pre-processed EIT images and the moments of the profiles that are used for the classification of off-limb structures.
In Section~\ref{s:classif} we describe the training data set which is used to train a Support Vector Machine, and the results from the classification.
Section~\ref{s:detect} presents the application of this classification algorithm to the automatic detection of prominences and { Section \ref{s:reconstruct} discusses the reconstruction of these features by morphological image processing techniques.}
Finally, we discuss our results in Section~\ref{s:discuss}.

\section{Radial Intensity Profiles} \label{s:profiles}

EUV images at 304 \AA\ reveal prominences as structures of enhanced brightness compared to
the surrounding quiet corona. The automatic detection of prominences in such an image is
however greatly complicated by the frequent presence in the same image of active regions (ARs),
where emission is also enhanced compared to quiet Sun areas.
There are some differences in the morphology and appearance of prominences and ARs: the latter
are typically brighter and more compact. In addition, the temperature at which ARs emit 
is different than for prominences, and therefore the response of the instrument should 
also be different for these structures.

Our detection method aims to characterise differences between prominences and other types of regions
(either ARs or quiet corona) by analysing radial intensity profiles and calculating their moments.
Although radial intensity profiles for prominences and ARs may appear similar to the eye, 
actual extraction of these profiles shows that they are not, and their moments allow us
to capture their differences.

To extract the radial profiles, we calibrate the SOHO/EIT images using the standard \url{eit_prep} procedure
and transform to polar coordinates $(r,\theta)$ where $r$ indicates the radial coordinate (distance
from the centre of the solar disk as specified in the image's FITS headers) and $\theta$ the angular coordinate.
We then obtain a radial cut at a given angular location  $\theta_0$ by extracting all points with angular 
coordinate within $\pm$1$^{\circ}$
of $\theta_0$ and with radial coordinate $r$ satisfying $1.01<r<1.35~R_{\odot}$.

\subsection{Example Radial Intensity Profiles} \label{sec.examples}

\begin{figure}
	\begin{center}
         \includegraphics[width=1.0\textwidth]{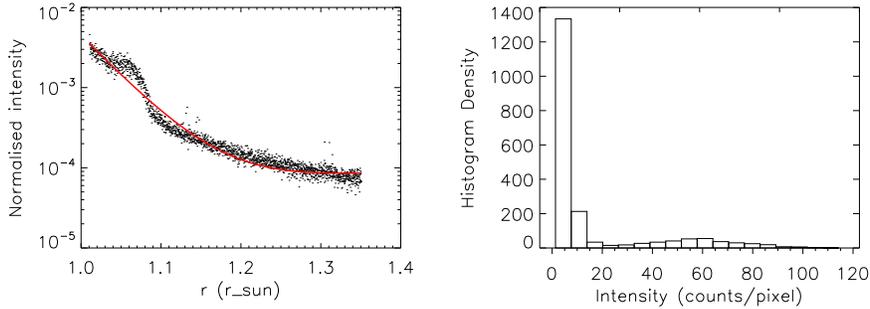}
	\end{center}
	\caption{Radial profiles and intensity histogram for a prominence.}
	\label{prom_profiles}
\end{figure}

A typical radial intensity profile for a prominence is shown in Figure~\ref{prom_profiles}{. The left panel of the figure shows the intensity profile} normalised to the sum of all intensity values in the profile. 
{A} double power law is fitted to the data points (solid line). The right panel of Figure~\ref{prom_profiles} shows the intensity histogram for all points in the radial profile.

\begin{figure}
	\begin{center}
         \includegraphics[width=1.0\textwidth]{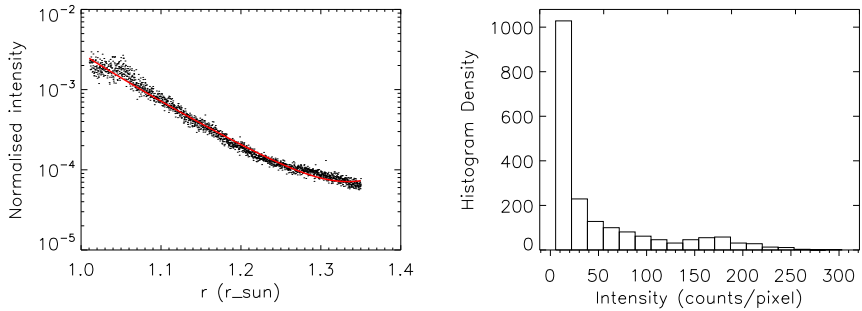}
	\end{center}
	\caption{Radial profiles and intensity histogram for an active region.}
	\label{ar_profiles}
\end{figure}

Figure~\ref{ar_profiles} shows radial intensity profiles and the intensity histogram for an AR, with the same format as in Figure~\ref{prom_profiles}.  
Comparing Figures~\ref{prom_profiles} and \ref{ar_profiles}, one notices that the prominence profile displays a rather sharp boundary at about 1.08~$R_{\odot}$, while emission from the AR gives a smoother curve. The comparison of the intensity histograms shows that, as one would expect, AR emission is typically characterised by higher intensities.

\begin{figure}
	\begin{center}
         \includegraphics[width=1.0\textwidth]{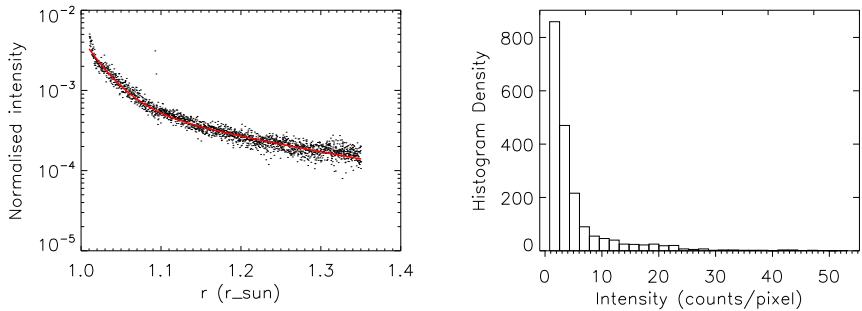}
	\end{center}
	\caption{Radial profiles and intensity histogram for a quiet corona region.}
	\label{quiet_profiles}
\end{figure}

Figure~\ref{quiet_profiles} shows the radial intensity profile for the quiet corona and the corresponding intensity histogram, characterised by much lower intensities than for the prominence and AR examples, as well as a different variation with $r$.

\subsection{Moments}

Moments represent a powerful way to capture the features of the radial intensity profiles 
and intensity histograms described in Section \ref{sec.examples}. We experimented with a variety of possible definitions of moments, and identified two sets that allow us to best discriminate between different limb structures. 

A first set of moments is obtained from the intensity histograms of the profiles as follows:
we define as $y_k$ the starting location of the $k$-th intensity bin in the histogram and as 
$n_k$ the number of pixels in the radial profile with intensity falling in the $k$-th bin. 
The following set of six moments is then introduced:
\begin{equation} \label{eq.m1}
	\mu_1^m = \frac{1}{N} \sum_{k}  (y_k-a)^m n_k \quad,\quad m=0,...,5
\end{equation}
with
\begin{equation} \label{eq.m1a}
	a =  \frac{\sum_k{y_k n_k}}{\sum_k{n_k}}
\end{equation}
where $a$ is the average intensity and $N$ the total number of pixels in the profile.

The second set of moments is calculated from the radial intensity profiles
by means of the following procedure: we indicate as $r_i$ the radial 
position of $i$-th data point in the profile,
as $y_i$ its intensity, and we define normalised intensities (as displayed in the {left} panels of 
Figures~\ref{prom_profiles}\,--\,\ref{quiet_profiles}) as 
\begin{equation}
      y_{Ni}   =  \frac{ y_i}{\sum_i{y_i}} \ .
\end{equation}
We fit each normalised intensity profile with a double power law 
(solid line in Figures~\ref{prom_profiles}\,--\,\ref{quiet_profiles}) defined by:
\begin{equation}
      y_{fit}(r) = a_0 \, r^{a_1} + a_2 \, r^{a_3} \ .
\end{equation} 
We can then obtain a new variable $z_i$ = $| y_{Ni}-y_{fit}(r_i)|$ and calculate its moments, defined by:
\begin{equation} \label{eq.m2}
	\mu_2^m = \frac{1}{N} \, \sum_{i}  (r_i-b)^m z_i \quad,\quad m=0,...,5
\end{equation}
with
\begin{equation} \label{eq.m2b}
	b =  \frac{\sum_i{r_i z_i}}{\sum_i{z_i}}
\end{equation}
This provides a second set of six moments.

\section{Classification} \label{s:classif}

Our aim is to obtain a decision rule that will allow us to classify a given radial
profile as belonging to one of three classes, either ``Prominence'', ``AR'', or ``Other'',
depending on the values of its moments.
Once a given profile has been identified as of prominence type, then it will be possible
to retrieve the entire prominence by means of a region-growing technique.

\subsection{Training Data} \label{s:training}

To obtain the required decision rule, we first assemble a dataset of
EIT images that will be used as training data for the algorithm.
We select a set of 25 SOHO/EIT 304~\AA\ images consisting of
approximately one image every six months between 1996 and 2006, covering a full solar cycle.
 
We extract ten radial profiles per image, for a total of 250 radial profiles,
that we classify by eye as either ``Prominence'', ``AR'', or ``Other''.
We also use visual inspection of SOHO/EIT images at 171~\AA\ and 195~\AA\ to distinguish more clearly between cool and hot structures. In principle, prominences will appear bright in the 304~\AA\ channel, and will be dark or invisible in the other two channels. 
Active regions will appear bright in all three EUV channels.

We randomly choose 167 profiles as the training set and feed
the other 83 to our classification algorithm as unlabelled profiles in order to test its performance.

\subsection{Support Vector Machine (SVM)}

Simple linear classifiers plot data samples in a multidimensional feature space and seek to classify them by positioning a linear hyper-plane such that it partitions them into two different groups. In practice there can be many hyper-planes that partition the data and some of these may require only a small perturbation in some of the data samples to lead to misclassification. 

Support Vector Machines, or SVMs \cite{cristianini}, are more sophisticated in that they partition the data using two parallel hyper-planes and adjust these so that the planes are the maximum distance apart without misclassifying any of the data samples. In practice it may be that it is not possible to separate a given set of data samples with a linear plane. In these cases a number of extensions exist which allow the SVM to deal with this situation. One approach is the so called \textit{soft classifier} approach \cite{1995Cortes} which allows a limited number of samples to be misclassified and seeks to minimise this number. Another approach is the so called \textit{kernel trick} \cite{1964Aizerman} that maps data into a higher-dimensional space in which it is linearly separable. 

SVMs represent a powerful technique for general classification, regression, and outlier detection with an intuitive model representation. For binary classification, we look for the optimal separating hyperplane between two classes by maximizing the margin  between the classes' closest points. The points lying on the boundaries are support vectors, and the middle of the margin is the optimal separating hyperplane. When we cannot find a linear separator, data points are projected into a higher-dimensional space where they effectively become linearly separable. The whole task can be formulated as a quadratic optimization problem which can be solved by known techniques. A program able to perform all these tasks is called a Support Vector Machine.

In this work we use the \texttt{libsvm} C++ implementation of SVMs \cite{libsvm} through an interface using the R language (\url{http://www.stats.bris.ac.uk/R/web/packages/e1071/}). The kernel used in the algorithm for training and prediction is a radial basis function, parametrized by one constant. Therefore we have two free parameters for training the SVM. The other parameter is a penalizing parameter for constraints violation (points lying between the margins). The choice of the parameters is crucial for obtaining good results, and we have therefore conducted trials over a large grid-space to find the best results.

\subsection{SVM results} \label{s:results}

Table \ref{svmresults1} displays the results of the SVM analysis using the first set of moments as defined in Equations.~(\ref{eq.m1})\,--\,(\ref{eq.m1a}).
\begin{table}
	\caption{ SVM results for the first set of moments.}
	\label{svmresults1}
	\begin{tabular}{lccc}
		\hline                   % horizontal line
		& AR (true)	& Other (true)	& Prominence (true) \\
		\hline
		AR (predicted)		& 21	& 0	& 4  \\
		Other (predicted)	& 0	& 22	& 5  \\
		Prominence (predicted)	& 0	& 1	& 30 \\
		\hline
	\end{tabular}
\end{table}
The true number of prominences, as determined by our visual identification, is 39. Of these, 30 are correctly identified by our algorithm. 
The true number of ARs is 21 and the algorithm correctly identifies all of them.
The remaining profiles, corresponding to the quiet corona, are 23 in number, 22 of them are correctly classified by the SVM algorithm.
The total number of misclassified profiles is ten, giving a misclassification probability of 12\%.

In order to improve the success of the algorithm, \textit{i.e.} to decrease the misclassification probability, we now use the second set of moments defined in Equations.~(\ref{eq.m2})\,--\,(\ref{eq.m2b}) in combination with the previous one. The SVM algorithm is now trained with 12 moments describing each radial profile.
Table \ref{svmresults2} displays the results of the SVM analysis using the two sets of moments. Note that the 83 profiles which are used for testing the algorithm are not the same as for Table~\ref{svmresults1}, since they are chosen randomly. This is the reason why the number of \textit{true} profiles in each category is not the same as in Table~\ref{svmresults1}.
\begin{table}
	\caption{SVM results for the two sets of moments.}
	\label{svmresults2}
	\begin{tabular}{lccc}
		\hline                   % horizontal line
		& AR (true)	& Other (true)	& Prominence (true) \\
		\hline
		AR (predicted)		& 19	& 0	& 0  \\
		Other (predicted)	& 0	& 24	& 1  \\
		Prominence (predicted)	& 2	& 3	& 34 \\
		\hline
	\end{tabular}
\end{table}
The true number of prominences, as determined by our visual identification, is 35, of which 34 are correctly identified by the algorithm.  
The true number of ARs is 21 and the algorithm correctly identifies 19 of them.
24 out of 27 remaining profiles corresponding to the quiet corona are correctly classified.
The total number of misclassified profiles is now six, giving a misclassification probability of 7\%.

\begin{figure}
	\begin{center}
         \includegraphics[width=1.0\textwidth]{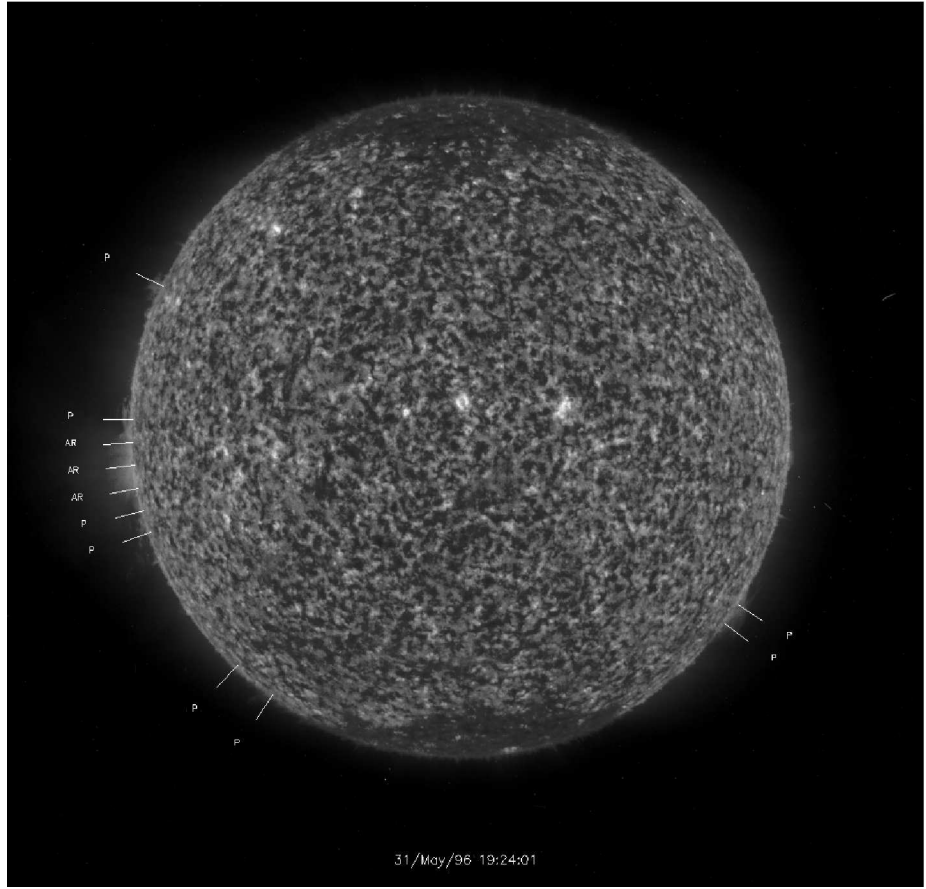}
	\end{center}
	\caption{Full-Sun EIT image at 304~\AA\ on 31 May 1996 showing detected prominences (P) and active regions (AR).}
	\label{fig:960531}
\end{figure}
\begin{figure}
	\begin{center}
         \includegraphics[width=1.0\textwidth]{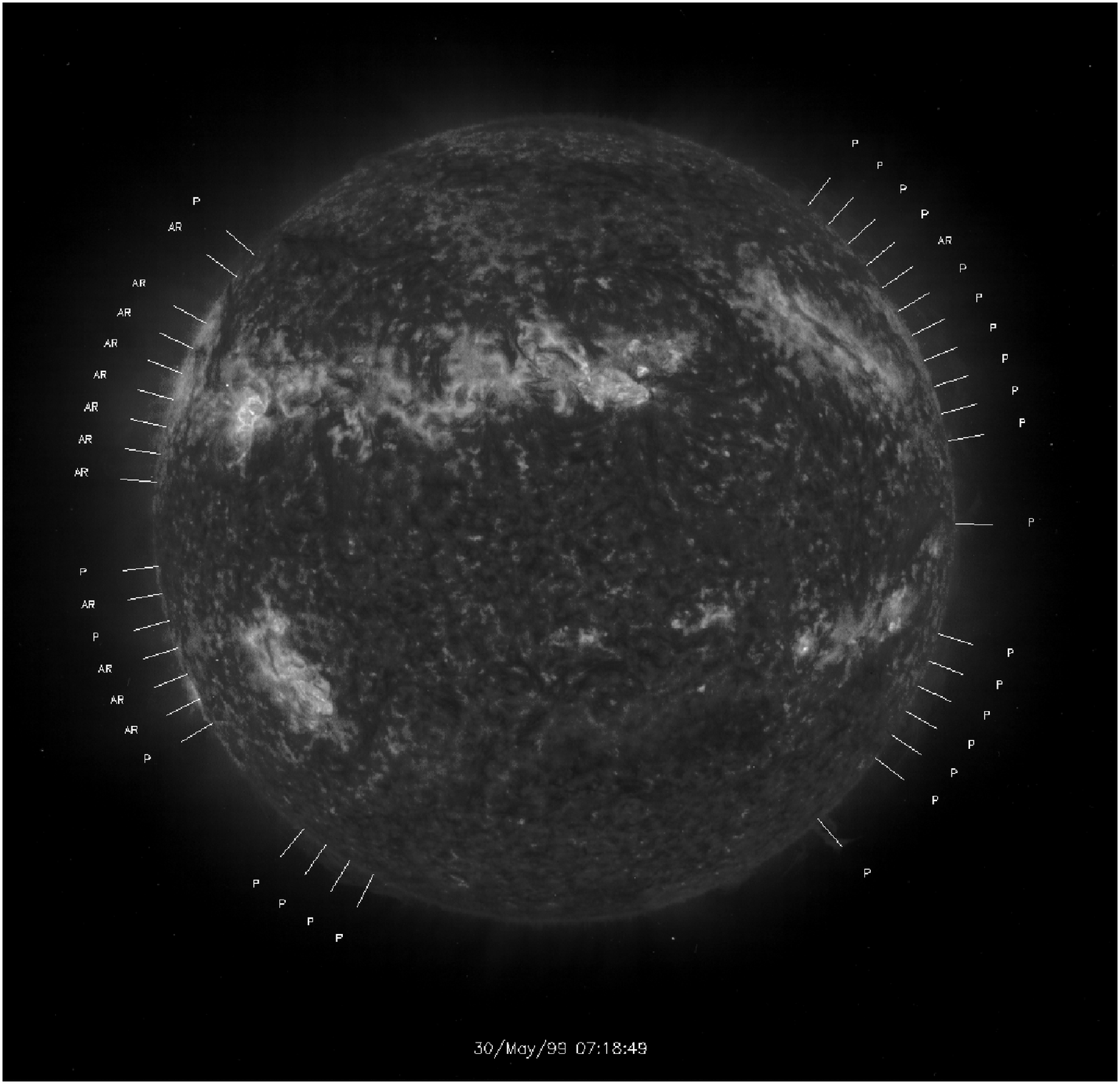}
	\end{center}
	\caption{Full-Sun EIT image at 304~\AA\ on 30 May 1999 showing detected prominences (P) and active regions (AR).}
	\label{fig:990530}
\end{figure}

\section{Detection of Limb Prominences} \label{s:detect}

{ The steps required}
to create a catalogue of solar prominences from the EIT database of observations at 304~\AA, { are as follows}:
\begin{enumerate}
	\item Given a new EUV image, we select a number of angles (up to 90 to have a sufficient coverage of the limb) at which we compute the radial intensity profiles and the associated moments.
	\item We use the now-trained SVM algorithm to automatically decide what type of structure is present.
	\item Where prominences are detected, we apply growing region techniques to recover the whole structure.
	\item We calculate barycentre, area, and angular / radial extension of the whole structures found on the image.
	\item We proceed to a subsequent image. A comparison between subsequent images enables us to track any changes in time for a given prominence, such as disappearance due to the passage on the disk or behind the limb, eruption, or other types of disappearance.
\end{enumerate}
We illustrate the first two points above.
We now consider all 25 images that were used to manually label the intensity profiles as described in Section~\ref{s:training}. For each image we  construct 90 profiles equally spaced by four degrees. We then calculate the 12 moments of the intensity profiles as discussed in Section~\ref{s:results}. 
Two examples are shown in Figures~\ref{fig:960531} (close to solar minimum) and \ref{fig:990530} (close to solar maximum).
Figure~\ref{fig:960531} shows that all prominences are detected. In addition, an active region on the east limb is also detected. Prominences are also detected on the southeast limb.  
Figure~\ref{fig:990530} shows a full-Sun image closer to the maximum of the solar-activity cycle. It is obvious that the background diffuse emission of the corona is brighter than earlier in the solar cycle (Figure~\ref{fig:960531}). Despite the lower contrast between prominences and the corona, all visible prominences are detected as such. All active region loop systems are detected as such. There are a couple of less clear cases of prominence or AR detection, that might be discarded at the stage of catalogue production because they do not meet the criteria for inclusion (i.e. they will not meet the minimum size requirement).

{It takes about 17 seconds to apply the standard SSW corrections on an EIT image, subsequently take 90 radial profiles, and classify them using the SVM algorithm. We expect to be able to reduce this  time thanks to various optimisations in the codes. Also, we will aim to reduce the number of profiles (less than 90) while still keeping a good coverage of the solar limb.}

In the next Section we apply region-growing techniques to reconstruct the whole prominence once a detection is made.
\begin{figure}
	\begin{center}
         \includegraphics[width=1.0\textwidth]{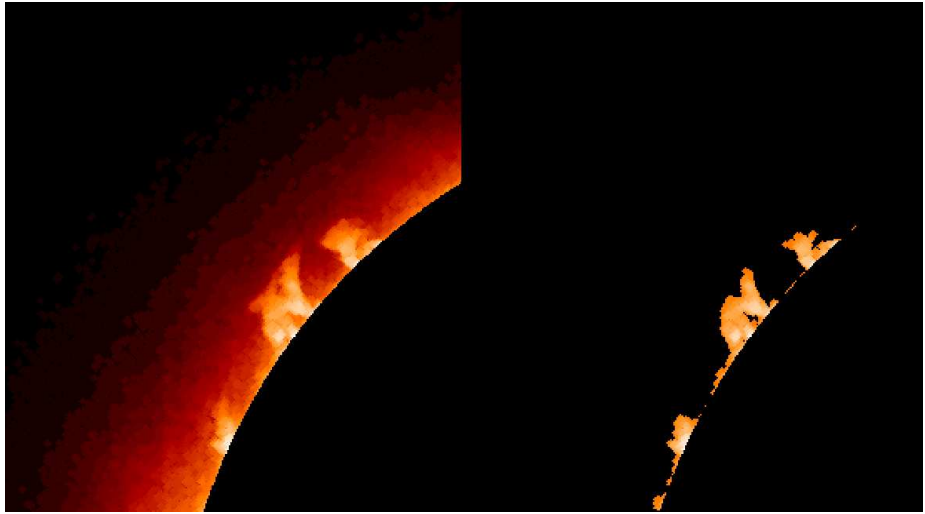}
	\end{center}
	\caption{Example of morphological opening by reconstruction for a prominence observed by SOHO EIT on May 26 2006. The left panel shows the subfield of the original image on which the reconstruction is run. The right panel shows the output.}
	\label{fig:reconstruct}
\end{figure}

\section{Prominence Reconstruction} \label{s:reconstruct}

{ Once our detection procedure has identified that a prominence exists at a particular angular location, we automatically reconstruct the entire prominence using morphological image processing. This allows us to obtain the pixels that belong to the prominence and extract its characteristics in the next step.}

{ We start by applying morphological opening to the image to remove spurious bright pixels (e.g. cosmic ray hits).
To start the reconstruction, we use the information from the detection algorithm to identify a starting pixel (the marker) that belongs to prominence. Next we apply morphological opening by reconstruction using this marker and a circular structure element with radius of 10 pixels.}

{ Figure \ref{fig:reconstruct} shows an example of prominence reconstruction obtained using our method. The pixels belonging to this complex prominence are identified. The size of the subimage shown is 277x306 pixels and it is likely that the three separate regions identified belong to the same prominence.}

\section{Discussion and Conclusions} \label{s:discuss}

In this paper we present the first attempt to automatically detect solar prominences above the limb using only one type of observation, namely EUV images at 304~\AA. 
Despite the fact that prominences and other coronal off-limb structures such as active regions sometimes look similar on standard EIT images, the method of the moments of the intensity profiles, coupled with a Support Vector Machine for the classification of the moments in three categories, is working and yields  a small rate of misclassification.
While our initial aim was to detect and construct a catalogue of limb prominences from the 304~\AA\ images produced by EIT \,--\, and also from STEREO/EUVI \,--\, it appears that this method is also effective in identifying active region loop systems above the limb.

The biggest challenge in this exercise is to remove the effect of the background corona. The diffuse emission that we observe is a mixture of true emission (as we can observe that it is brighter around the solar maximum than at solar minimum) and of the instrument's scattered light. At present there is no automatic procedure to remove the latter component from the images, but observations performed during the Mercury transit of 15 November 1999 may help \cite{2004SoPh..219..217A}. Removing the instrumental component of the scattered light will increase the contrast between the background corona and the other structures.

We plan to run our detection algorithm on the entire SOHO/EIT dataset of 304~\AA\ images to generate a 
catalogue of limb prominences, to be made publicly available to the scientific community via a
web page and via Virtual Observatory interfaces developed by the AstroGrid project.
We will also be implementing our detection method on 304 \AA\ images from STEREO/EUVI and SDO/AIA
for the generation of a limb prominence catalogue for these missions.
The resulting catalogue can be used to perform a statistical study of, \textit{e.g.}, the large-scale properties of prominences as seen on the limb over several solar cycles (thanks to the continuity between EIT, EUVI, and AIA observations). The relationship between limb prominences and CMEs will also be investigated.

%%%%%%%%%%%%%%%%%%%%%%%%%%%%%%%%%%%%%%%%%%%%%%%%%%%%%%%%%%%%%%%%%%%%%%%%%%%
%% Acknowledgements
%
\begin{acks}
	We acknowledge support from the AstroGrid project as part of the 2008 Science Tool Call.
	NL thanks the Royal Society for its financial support through grant CG080266.
We acknowledge support from ISSI through funding for the International Team on SDO data mining and exploitation in Europe.
	This research has made use of NASA's Astrophysics Data System.
\end{acks}

%%% %%%%%%%%%%%%%%%%%%%%%%%%%%%%%%%%%%%%%%%%%%%%%%%%%%%%%%%%%%%
%% Bibliography
%
% Using BibTeX
%

\bibliographystyle{spr-mp-sola}

\bibliography{improc}  

\end{article} 
\end{document}